\documentstyle[aps,prl,epsf,twocolumn]{revtex}
\begin{document}
\def\bea{\begin{eqnarray}}
\def\eea{\end{eqnarray}}
\def\a{\alpha}
\def\d{\delta}
\def\p{\partial} 
\def\nn{\nonumber}
\def\r{\rho}
\def\rv{\bar{r}}
\def\la{\langle}
\def\ra{\rangle}
\def\e{\epsilon}
\def\o{\omega}
\def\n{\eta}
\def\g{\gamma}
\def\break#1{\pagebreak \vspace*{#1}}
\def\f{\frac}
\draft
\title{Topological Phases near a Triple Degeneracy}
\author{Joseph Samuel and Abhishek Dhar}
\address{ Raman Research Institute,
Bangalore 560080\\ }
\date{\today}
\maketitle
\widetext
\begin{abstract}
We study the pattern of three state topological phases that appear
in systems with real Hamiltonians and wave functions. We give a simple
geometric construction for representing these phases.  We then 
apply our results to understand previous work on three state phases.
We point out that the ``mirror symmetry'' of wave functions 
noticed in microwave experiments can be simply understood in our framework.
\end{abstract}

\narrowtext

There has been considerable work on the Geometric phase in recent
years~\cite{berry,shapere,herzberg}, 
and the phenomenon is now quite well understood in differential geometric
terms as the twisting of a line bundle over the ray space of quantum mechanics.
In a two state system,
this abstract picture can be
made quite visualisable using the Poincare sphere. The ray space
is identified with the Poincare sphere  and the geometric phase 
is half the
solid angle enclosed by the ray in its quantum evolution. Thus, for two
state systems, the abstract differential geometric picture can be made quite
concrete and brought to bear on the interpretation of experiments. Indeed,
this field has seen a fruitful interplay between theory and experiment. 
The geometric phase appears in a wide variety of wave phenomena.
The theory has been tested in situations as diverse as optics \cite{shapere},
molecular spectroscopy \cite{mead}, nuclear magnetic resonance
\cite{shapere} and microwave cavities \cite{lauber}. 

For three state systems the situation is not as satisfactory. While
the abstract differential geometric picture is of course, correct and complete, 
it does not lead to simple visualisable pictures or formulae that can
directly be compared with experiment. There have been attempts in this
direction~\cite{Arvind}, 
but it is fair to say that the general three state Geometric
phase is considerably less accessible to the experimental community than
the two state geometric phase.
In this paper we take up a special case of
this general problem: our Hamiltonians (and wave functions) are restricted
to be real. 
This restriction considerably simplifies the problem and 
renders it tractable.  While complex Berry phases take values on the
unit circle and are correctly described as geometric, real Berry phases
can only take values ${\pm 1}$. These should be thought of as topological
phases, since they are insensitive to continuous deformations of the 
system history.  Even though these real Berry phases are topological,
the pattern of three state topological  phases that emerges is surprisingly rich, as we will see.
The aim of this Letter is to explore this rich structure and present it
in a form suitable for direct comparison with experiments.

The motivation for this work \break{1.2in}
comes from 
experiments with 
microwave cavities. 
Lauber et al~\cite{lauber} have performed experiments
in which they perturb a rectangular microwave cavity by deforming its shape
adiabatically.
The unperturbed cavity has a three fold degeneracy, which is lifted by the
perturbation. As the perturbation is varied cyclically, the system returns to
its original state but sometimes picks up a topological phase of $\pi$.  
There has been some discussion of the interpretation of this experiment.
Lauber et al suggested that the phase changes they were seeing were not
related to Berry's phase but were a new and independent phenomenon
related to ``mirror symmetry''. Manolopoulos
and Child~\cite{Mano} posed a specific theoretical problem and partially solved
it, thereby gaining insight into the pattern of Berry phases that are expected
from the general theory.  Ref.~\cite{Mano} also applied their theory to the 
experiment of Lauber et al and were able to explain the experimentally 
observed closed path phases
as Berry phases.  This work has been criticised by Pistolesi and
Manini~\cite{Pistol} on the grounds that the formalism of
Ref.~\cite{Mano} incorrectly  
predicts the
open path \cite{sam} phases. They attribute this to the presence of additional
``satellite'' degeneracies near the main degeneracy which were not taken
into account in Ref.~\cite{Mano}. 

In this paper we will completely solve the problem posed in Ref.~\cite{Mano}.
We first motivate and pose the problem 
that we address. We then present our solution to this problem. Finally
we use our solution to critically understand the previous theoretical
and experimental work in this area.

A general perturbation of the system will remove the three-fold degeneracy
and can be represented in the adiabatic approximation by a $3\times 3$ symmetric
matrix, $H$.
(The approximation consists of neglecting transition amplitudes
between the three states of interest and other states.) We are interested in
the eigenspaces of $H$. These are not affected by rescaling $H$ or adding
a multiple of the identity to it. 
One can thus arrange that $H$ be traceless $Tr(H)=0$. The space of traceless symmetric
matrices is a five dimensional vector space. This space has a natural inner
product $\la H_1,H_2 \ra=Tr(H_1H_2)$. Let us choose a basis ${Q_\alpha, 
\alpha=0,1,2,3,4}$ in this vector space which is orthonormal:
 $\la Q_\alpha, Q_\beta \ra =\delta_{\alpha \beta}$.
We can expand $H$ in this basis $H=\Sigma_{\alpha} x^{\alpha} Q_{\alpha}$.
It is convenient to normalise $H$ by the condition $\la H,H \ra =1$. This
results in the $x^{\alpha}$ satisfying $\Sigma_{\alpha} x^{\alpha} x^{\alpha}=1$,
which describes a sphere $S^4$ in five dimensions.

At some points of $S^4$, $H$ has doubly degenerate levels and these points
are said to belong to $\cal D$. Let us arrange the eigenvalues in decreasing
order so that $\lambda_1\ge \lambda_2 \ge \lambda_3$, the equalities holding
only at points of $\cal D$. If $\lambda_1=\lambda_2>\lambda_3$, we say that
the point belongs to ${\cal D}^+$ and if  $\lambda_1>\lambda_2=\lambda_3$
we say that the point belongs to ${\cal D}^-$. The eigenvalues of $H$ satisfy
two identities 
$\lambda_1+\lambda_2+\lambda_3=0$ and $\lambda_1^2+\lambda_2^2+\lambda_3^2=1$
(from tracelessness and normalisation of $H$) and can be parametrised by 
an angle $\psi, 0\le\psi\le\pi/3$: $\lambda_1=\sqrt{2/3}\cos\psi, 
\lambda_2=\sqrt{2/3}\cos(\psi-2\pi/3),\lambda_3=\sqrt{2/3}\cos(\psi+2\pi/3)$.
Note that $\psi=0$ at points of ${\cal D}^-$ and
$\psi=\pi/3$ at points of ${\cal D}^+$.

Given a closed curve in the space of perturbations (which before normalisation
is five dimensional) which does not pass through points of $\cal D$, the standard
Berry phase lore would instruct us to diagonalise $H$ along the curve and
transport its eigenvectors by continuity to compute the phase change 
as the system is cyclically perturbed. The general problem of computing
the phase is greatly simplified if one restricts attention to
great circles on $S^4$, which is well-motivated
experimentally. 
In an actual
experiment~\cite{lauber}, one does not always explore the full five
dimensional parameter space. 
One needs to vary at least two parameters to 
effect a cyclic evolution and see a Berry phase. Following
ref.\cite{Mano}, we will restrict 
ourselves to a two dimensional parameter space $(x,y)$, chosen so that
$(0,0)$ represents the triply degenerate Hamiltonian $H=0$. Expanding
the Hamiltonian $H(x,y)$ in a Taylor expansion about the degenerate point
we find
\begin{equation}
H(x,y)=f x+g y + O(x^2,y^2),
\end{equation}
where $f=\frac{\partial H}{\partial x}(0,0)$ and 
$g=\frac{\partial H}{\partial y}(0,0)$. Thus the perturbations 
span a two dimensional plane in the space
of $3\times3$ traceless symmetric
matrices. The intersection of this plane with the sphere $S^4$ is a great
circle (or geodesic) on $S^4$.  The great circle can be specified by giving
two (non antipodal) points on it. We choose $(F,G)$ an orthonormal
basis in the $(f,g)$ plane and write:
\begin{equation}
H(\theta)=\cos \theta F+\sin \theta G
\label{Hoftheta}
\end{equation}
where $\theta$ varies from $0$ to $2\pi$ and $ \la F,G \ra =0$.
If $H(\theta)$ is nondegenerate for all $\theta$, one can uniquely follow the
eigenvectors of $H(\theta)$ as a function of $\theta$ and the adiabatic
Berry phase $\gamma_i$ of the state $|i>$ is well defined. 
From \cite{Mano}  it follows that $\gamma_2=1$ and $\gamma_1=\gamma_3=\gamma=\pm1$.
Without loss of generality, 
we can suppose that $F$ is in diagonal form: 
$F=\lambda_1 |1><1|+\lambda_2|2><2|+
\lambda_3 |3><3|$, since this can be arranged by using the eigenvectors
of $F$ as an orthonormal basis. The problem posed in ref.
\cite{Mano} is: given the pair $(F,G)$ compute the topological phase $\gamma$ 
for the great circle $H(\theta)$. This problem has been solved in
ref.~\cite{Mano}
in the case where $G$ is of a special ``bipartite'' form. Another special case
where the answer is known is when $F$ and $G$ share a common eigenvector. By
projecting orthogonal to this eigenvector, one can reduce this case to the 
two-state topological phase. Then the projection of $H$ can be expanded in terms of the
Pauli matrices as $\cos \theta \sigma_z
+\sin \theta \sigma_x$ and its eigenvectors clearly reverse sign as $\theta$
goes from $0$ to $2 \pi$.  For a general pair $(F,G)$ there does not 
at present exist a simple rule to determine the topological phase.  This
is the purpose of this Letter.
We give the general solution to this problem below.

Regard $F$ as the ``north pole'' of $S^4$. Geodesics through $F$ are characterised
by $G\in S^3$, where $G$ lies in the ``equator'' of $S^4$. We will first
locate all points in $S^3$ which lead to geodesics passing through ${\cal D}$.
We refer to these as ``degenerate'' points and they form a set ${\cal
B}\in S^3$. 
Regarded as a function of $G$, the topological phase can only
change when $G$ passes through ${\cal B}$, when the phase becomes ill-defined.
Thus ${\cal B}$ divides $S^3$ into regions, each of which has the same topological
phase. Our next step is to locate the set ${\cal B}$ in $S^3$ and thus split
up $S^3$ into regions of constant topological phase.

We first characterise the degenerate set $\cal D$ in $S^4$. Clearly, 
$\cal D=\cal D^+
\cup \cal D^-$, which are disjoint sets. The points of 
${\cal D}^{\pm}$ can be 
written as $D^{\pm}_{\hat n}=\pm\f{1}{\sqrt{6}}(1-3 |\hat n><\hat n|)$,
where $|\hat n>$ is some unit ket vector (the nondegenerate
eigenvector of $D$). Since $|\hat n>$ and $-|\hat
n>$ define the same $D^{\pm}_{\hat n}$, ${\cal D}$ is the disjoint 
union of two $I\!\!R P^2$ s.
To find the degenerate set ${\cal B}$ 
is easy:  the plane containing the vectors $F$ and $D^{\pm}_{\hat n}$ 
intersects the sphere $S^4$ in a great circle. The intersection of this
great circle with the equator of $S^4$ is simply the projection of 
$D^{\pm}_{\hat n}$  orthogonal to $F$, suitably normalised:
\begin{equation}
B^{\pm}_{\hat n}=\frac{D^{\pm}_{\hat n}- (D^{\pm}_{\hat n},F) F}
{[(D^{\pm}_{\hat n},D^{\pm}_{\hat n})
-(D^{\pm}_{\hat n},F)^2]^{1/2}}
\end{equation}
When $G$ is equal to any of these points $B^{\pm}_{\hat n}$, the
topological phase is ill defined. These are the degenerate points in
$S^3$. 
Let us compute their co-ordinates 
explicitly. Choose matrices $Q_{\alpha}$ (similar parametrisations appear 
in \cite{jahn,panch,sadun})
\bea
&& Q_0=\f{1}{\sqrt{6}}{\rm diag}(2,-1,-1), \nn \\
&& Q_1=\f{1}{\sqrt{2}}\left( \begin{array}{ccc} 0 & 0 & 0  
\\ 0 & 0 & 1 \\ 0 & 1 & 0
\end{array} \right),~
Q_2=\f{1}{\sqrt{2}}\left( \begin{array}{ccc} 0 & 0 & 1  \\ 0 & 0 & 0 \\ 1 & 0 & 0
\end{array} \right), \nn \\ 
&& Q_3=\f{1}{\sqrt{2}}\left( \begin{array}{ccc} 0 & 1 & 0  \\ 1 & 0 & 0 \\ 0 & 0 & 0
\end{array} \right),~Q_4=\f{1}{\sqrt{2}} {\rm diag}(0,1,-1). \nn
\eea
F is diagonal and therefore a linear combination of $Q_0$ and $Q_4$:
$F=\cos\psi Q_0+\sin \psi Q_4$. 
Let us define $E=-\sin\psi Q_0+\cos \psi Q_4$ a diagonal matrix
orthogonal to $F$. In the basis $(F,E, Q_1,Q_2,Q_3)$, we find that
$G=g_4 E+g_1Q_1+g_2 Q_2+g_3 Q_3$,
where the $g$'s satisfy $g_1^2+g_2^2+g_3^2+g_4^2=1$, which describes
$S^3$. Similarly expanding the degenerate points $B^{\pm}_{\hat n}$ in this
basis $B^{\pm}_{\hat n}=b_4 E+b_1Q_1+b_2 Q_2+b_3 Q_3$, we find
by explicit computation
\begin{eqnarray}
&& b_1=\mp\sqrt{3} \frac{n_2 n_3}{\sqrt{1-c^2}}, ~  b_2=\mp\sqrt{3}
\frac{n_1 n_3}{\sqrt{1-c^2}} \nn \\
&& b_3=\mp\sqrt{3} \frac{n_2 n_1}{\sqrt{1-c^2}},~ b_4=\pm
\frac{s}{\sqrt{1-c^2}}  
\label{bad}
\end{eqnarray}
where $c=\sqrt{3/2}<\hat n|F|\hat n>$ and 
$s=-\sqrt{3/2}<\hat n|E|\hat n>$. Defining angles
$\psi_1=\psi,\psi_2=\psi-2\pi/3,\psi_3=\psi+2\pi/3$, $c$
and $s$ can be expressed as: $c=\sum_i \cos{\psi_i} n_i^2;~s=\sum_i
\sin{\psi_i} n_i^2$.     

Since $G$ and $-G$ represent the same geodesic oppositely traversed, it
is clear that they have the same topological phase.  Also,
reversing the sign of two of $(g_1,g_2,g_3)$  amounts to a $\pi$ rotation
about one of the principal axes of $F$. This is merely a similarity 
transformation of the pair $(F,G)$ and does not alter the topological phase.
By such phase preserving transformations one 
can arrange for $(g_1,g_2,g_3)$to be nonnegative.
($g_4$ can have either sign.) It is easily seen from (\ref{bad}) that 
points of $S^3$ where $(g_1,g_2)$ or $(g_2,g_3)$ vanish 
are degenerate points. Points of $S^3$
where $(g_1,g_3)$ vanish are \cite{foote} in $\cal B$  
if $g_4< g_4^c= 
{-[1-3/4 {\rm cosec}^2(\psi+\pi/3)]^{1/2}}$. If $g_4>g_4^c$, the
topological phase is well defined and
$F$ and $G$ have a common eigenvector $(0,1,0)$ 
and the topological phase is then $-1$. 
Points with two or more of $(g_1,g_2,g_3)$ vanishing
will be excluded from the following discussion.

We now introduce new coordinates $(v_1,v_2,v_3)$ given by
\begin{eqnarray}
&& v_1=A g_2^2g_3^2,~v_2=A g_3^2g_1^2,~  
v_3=A g_1^2g_2^2, \label{vcoord}
\\ 
&&{\rm{where}}~~~ A=(g_1^2 g_2^2+g_2^2 g_3^2+g_1^2 g_3^2)^{-1}. \nn
\end{eqnarray}
The $v$'s are positive and satisfy $v_1+v_2+v_3=1$, which 
defines a triangular region. 
Let us define the following Cartesian coordinates in the plane $v_1+v_2+v_3=1$
\begin{equation}
C=\Sigma_{i}v_i \cos \psi_i,~
S=\Sigma_{i} v_i\sin \psi_i.
\label{CScoord}
\end{equation}
Let $g_4$ be
held fixed. Then  $g_1^2+g_2^2+g_3^2=1-g_4^2=r^2$ and $(g_1,g_2,g_3)$
determine a point in the $C-S$ plane via (\ref{vcoord},\ref{CScoord}). As $(g_1,g_2,g_3)$
vary over permissible values, $(C,S)$ describe an equilateral triangle
$\Delta$. The orientation of this triangle relative to the $C$ axis 
is controlled by the eigenvalue
parameter $\psi$ of $F$. We now locate the degenerate points 
in $\Delta$. These points are given parametrically by Eq.~(\ref{bad}). The 
$v$ coordinates of these points are $v_1=n_1^2,v_2=n_2^2,v_3=n_3^2$. As 
$(n_1,n_2,n_3)$ range over permissible values $(n_1,n_2,n_3>0,n_1^2+n_2^2+n_3^2=1)$
$(C,S)$ describe all the points in $\Delta$. The degenerate points of
$\Delta$ 
are located by requiring that $b_4=g_4$. Using the last of
Eqs.~(\ref{bad}) it follows that these points describe an ellipse 
in the $C-S$ plane: $C^2+S^2/{(1-r^2)}=1$ with eccentricity determined
by $g_4$. The part of the ellipse inside $\Delta$ consists of two segments 
(Fig.~\ref{vplane}). The degenerate points consist of the lower segment
(for $g_4>0$)
and the upper segment  (for $g_4<0$). These clearly split up the 
triangle into two regions with constant topological phase. When $g_4=0$, 
the ellipse degenerates to a straight line parallel to the $C$ axis. 
As $g_4$ approaches $1$, the
ellipse expands to the unit circle. By slightly perturbing the matrix $G=Q_2$ so that
it is represented on the figure, we see  
that the  correct assignments are as shown in the figure. This figure and the 
algorithm given below for using it are the main results of this paper.

{\it The Algorithm:}
1. Given two $3\times3$ symmetric, perturbation 
matrices $(f,g)$, add a multiple of the identity to make them trace free and
choose linear combinations $(F,G)$ which are normalised and orthogonal:
$ Tr(F F)=Tr(G G)=1$, $Tr(F G)=0$.

2. Diagonalise F and write G in the basis in which F is
diagonal. Determine the eigenvalue parameter $\psi$ and construct
$E$. 

3. Expand $G$ in the basis $(E,Q_1,Q_2,Q_3)$ and determine
$(g_1,g_2,g_3,g_4)$. [eg. $g_2=Tr (G Q_2)$]. Apply appropriate
rotations so that $g_1,g_2,g_3$ are positive.
 
4. Draw an equilateral triangle $\Delta$ in the $(C,S)$ plane 
with vertices $(\cos{\psi_1},\sin{\psi_1})$, $(\cos{\psi_2},
\sin{\psi_2}
)$, $(\cos{\psi_3}, \sin{\psi_3})$, where $\psi_1=\psi$ and
$\psi_{2,3}=\psi \mp 2 \pi/3$.

5. Let $r^2=1-g_4^2$. Draw an ellipse $C^2+{S^2}/{(1-r^2)}=1$. The
sign of $g_4$ tells us which of the two segments of the ellipse 
contains the degenerate points.

6. Using Eqs.~(\ref{vcoord},\ref{CScoord}) locate 
$(g_1,g_2,g_3)$ in the $C-S$ plane and read off the topological phase from
the diagram: the phase is $-1$ for $R_1$, $-sign(g_4)$ in $R_2$ and $+1$
in $R_3$.

This completely solves 
the problem posed in \cite{Mano}.
Since our formalism involves a Taylor expansion of the Hamiltonian around the
triple degeneracy, we would expect the theory to agree with experiments that
explore a region of parameter space close to the degeneracy. If the perturbations
are such that the pair $(F,G)$ describe a degenerate point, then neither
the 
theory nor the experiment will produce  a definite answer for the topological
phase: the adiabatic approximation breaks down. 
Generically, for most shapes of cavities,
$G$ {\it will not be close to a degenerate point} and the first
order theory applies in a region around the triple degeneracy. 

For perturbed rectangular cavities 
the first order theory predicts a symmetry for the wave functions.
Let us start with the observation that the
unperturbed system has a discrete symmetry: reflection about a line normal to
the long side \cite{foota} and
bisecting the cavity ($P$).  The eigenstates of the unperturbed system
can be chosen to have a definite $P$ parity.
It can be easily seen \cite{comm} that the first order perturbations  can be 
decomposed into nonzero even ($F$) and odd ($G$) parts satisfying 
$Tr(FG)=0$.
Constructing $H(\theta)=\cos \theta F+\sin \theta G$, we see that 
$PH(\theta)P=H(2 \pi-\theta)$. Thus if $H(\theta) |\Psi_i(\theta)>=E_i(\theta) |\Psi_i(\theta)>$,
$|\Psi_i(2\pi-\theta)>=\sigma_i P|\Psi_i(\theta)>$, where $\sigma_i=\pm1$ is
constant by continuity. This relation between the wavefunctions at $\theta$
and $2\pi-\theta$ is what we refer to as ``mirror symmetry''.
Setting $\theta=0$ and using 
$|\Psi_i(2\pi)>=\gamma_i|\Psi_i(0)>$, where $\gamma_i$ is the topological 
phase, we find that $\sigma_i$ is the product of the topological phase 
and the parity of the $i$th state.

For the experiment of Lauber
et al \cite{lauber}, it turns out that
$F$ corresponds to $\psi=\cos^{-1}(23/26)$ and $G \approx
{(0.995,0.101,0,0)}$.  
$G$ is not actually a degenerate point but it is very close to one. The
point in $\cal B$ closest to $G$ is $({0.997, 0.066, 0.048,
-0.003})$. As a result,
the small splittings in the first order Hamiltonian
make the system very sensitive to second order perturbations. In fact these
higher order terms are able to perturb the Hamiltonian into $\cal D$, which
results in the appearance of two doubly degenerate ``satellites'' \cite{Pistol}.
As correctly pointed out in ref.\cite{Pistol}, these must also be taken
into
account to predict the ``open path'' Berry phases \cite{sam}. 
Fig.4 of \cite{Pistol}
shows the (shaded) regions in the parameter space where second order
perturbations are important. 
These correspond to $\theta\approx \pi/2,3\pi/2$
In the remaining regions first order theory
applies and we would expect the ``mirror symmetry'' described in the last 
paragraph to be present. For the three states with quantum
numbers $(n_x,n_y)=(7,1),(5,3),(2,4)$ the topological phase 
assignments are $\gamma_i=(-,+,-)$ and the parity assignments are 
$(+,+,-)$, which results in 
$\sigma_i=(-,+,+)$. The operation of reflecting the wave functions and
multiplying them by $\sigma_i$ relates the conjugate figures in \cite{lauber}. 
This symmetry was noticed by Lauber et al (see Fig. of \cite{lauber}) experimentally.
The mirror symmetry is particularly evident in the pairs 
(1,16), (2,15),(8,9),(7,10) where the first order
theory is expected to apply.
For other pairs (e.g the pairs (4,13),(5,12)) in
the vicinity of $\theta\approx \pi/2,3\pi/2$
the symmetry is only approximate as would be expected, since second order
effects are important. Thus the first order theory is able to account
for the experimentally observed approximate ``mirror symmetry'' in \cite{lauber}.

In summary, we have described a simple geometrical construction for
representing the topological phases of a three state system. The 
construction can be easily visualised and drawn on paper.  For this
reason it may serve as a useful tool to interpret experiments done
on three state topological phases. We have also given a theoretical
framework which is closely related to experimental investigations
of the topological phase.
We are also able to understand the ``mirror'' symmetry of wave functions 
seen in the experiments of Lauber et al. In fact the ``bipartite''
form of $G$ noticed by \cite{Mano} can be traced to this discrete
symmetry. Of course, our work also encompasses situations, where the 
unperturbed state has no discrete symmetries and $F$ and $G$ are general.
We believe this work will 
be useful to the community of physicists interested in the three state
topological phase.

{\it Acknowledgements:} It is a pleasure to thank Avinash Khare,
Vishwambhar Pati and Supurna Sinha for their comments on the manuscript.

\vbox{
\vspace{0.0cm}
\epsfxsize=6.0cm
\epsfysize=6.0cm
\epsffile{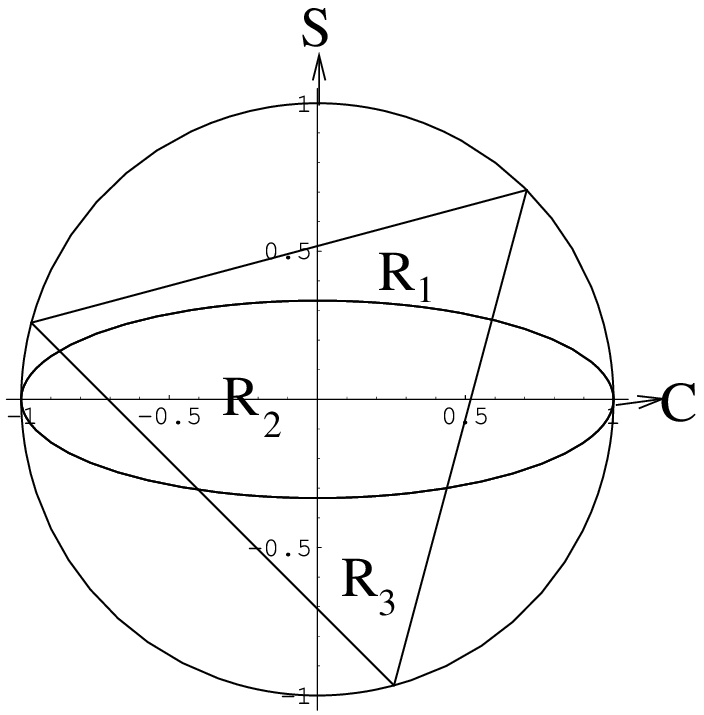}
\begin{figure}
\caption{Any point $(g_1,g_2,g_3)$ gets mapped to one of the three
regions $R_1$ $R_2$ and $R_3$ within the triangle. The topological
phase is $-1$ in $R_1$, $-sign{(g_4)}$ in $R_2$ and $+1$ in $R_3$. In
the figure, the ellipse and circle correspond to $g_4=1/3$ and $1$
respectively and the triangle to $\psi = \pi/4$.
\label{vplane} 
}
\end{figure}}


\begin{references}
\vspace{-1.5cm}
\bibitem{berry} M.V. Berry, Proc. Roy. Soc. {\bf A392}, 45 (1984).
\bibitem{shapere} A. Shapere and F. Wilczek, Geometric Phases in Physics
(World Scientific, Singapore, 1989).
\bibitem{herzberg} G. Herzberg and H.C. Longuet-Higgins, Discuss. Faraday Soc.
{\bf 35}, 77 (1963).
\bibitem{mead} C.A. Mead, Rev.Mod. Phys. {\bf 64}, 51 (1992).
\bibitem{lauber} H.-M. Lauber et al, Phys. Rev. Lett {\bf 72}, 1004 (1994).
\bibitem{Arvind} Arvind et al, J. Phys. A {\bf
30}, 2417 (1997); G. Khanna et al, Ann. Phys. (NY) {\bf 253}, 55 (1997);
M.S. Byrd et al, J. Math. Phys. {\bf 39}, 6125 (1998); B.C. Sanders et al
Phys. Rev. Lett. {\bf 86}, 369 (2001); E. Strahov, J. Math. Phys. {\bf
42}, 2008 (2001);
A. Alijah and M. Baer, J. Phys. Chem. {\bf A104}, 389 (2000).
\bibitem{Mano} D.E. Manolopoulos and M.S. Child, Phys. Rev. Lett {\bf 82},
2223 (1999).
\bibitem{Pistol} F. Pistolesi and N. Manini, Phys. Rev. Lett. {\bf
85}, 1585 (2000).
\bibitem{sam} J. Samuel and R. Bhandari, Phys. Rev. Lett. {\bf 60}, 2339 (1988).
\bibitem{jahn} H.A.  Jahn and E. Teller Proc. Roy. Soc. {\bf A161},
220 (1937);U. {\"O}pik and M.H.L. Pryce, Proc.R. Soc {\bf A238}, 425 (1957);
 W. Moffit and W.R. Thorson, Phys. Rev. Lett. {\bf 108}, 1251 (1957);
K.S. Mallesh et al, Pramana {\bf 49}, 371 (1997); E. Ercolesi et al,
Int. J. Mod. Phys. (To appear).
\bibitem{panch}
S. Pancharatnam, Proc. Roy. Soc. {\bf A330}, 271 (1972).
\bibitem{sadun} L. Sadun and J. Segert, J. Phys. A {\bf 22}, L111 (1989);
J.E. Avron et al, Phys. Rev. Lett. {\bf 61}, 1329 (1988);
Comm. Math. Phys. {\bf 124}, 595 (1989).
\bibitem{foote} This follows because for such values of $g_4$, we can find
$n_1,n_3$ ($n_2=0$ ) such that $G$ is the image of ${\hat n}$ under the
map (\ref{bad}).
\bibitem{foota}
Equivalently, one could reflect about a line normal to the short side;
the two operations are equivalent in the subspace.
\bibitem{comm} J. Samuel and A. Dhar, Submitted to Phys. Rev. Lett.
\end{references}
\end{document}